\documentstyle[aps,amssymb,epsfig,multicol]{revtex}
\def\be{\begin{equation}}
\def\ee{\end{equation}}
\def\bq{\begin{eqnarray}}
\def\eq{\end{eqnarray}}

 1

\begin{document}

\draft
\title{Band and filling controlled transitions in exactly 
solved electronic models}

\author{Fabrizio Dolcini and Arianna Montorsi
}
\address{Dipartimento di Fisica and Unit\`a INFM, Politecnico di Torino,
I-10129 Torino, Italy}
\date{\today}
\maketitle
\begin{abstract}
We describe a general method to study the ground state phase diagram of 
electronic models on chains whose extended Hubbard hamiltonian is formed 
by a generalized permutator plus a band-controlling term. The method, based
on the appropriate interpretation of Sutherland's species, yields
under described conditions a reduction of the effective Hilbert space. In
particular, we derive
the phase diagrams of two new models; the first one exhibits a
band-controlled insulator-superconductor transition at half-filling for 
the unusually high value $U_c=6 t$; the second one is characterized by a
filling-controlled metal-insulator transition between two finite
regions of the diagram.
\end{abstract}

\begin{multicols}{2}
Metal-Insulator and Insulator-Superconductor Transitions in 
chain systems of interacting electrons have recently become a 
matter of great interest for the physics of new compounds and 
devices\cite{REF}. Although many experimental data are nowadays 
at disposal, an important open question of this issue is to 
determine the Hamiltonian that could fairly describe these kinds 
of transitions. The task is particularly difficult just owing 
to the low dimensionality, which causes usual mean-field and 
perturbative approaches to often fail in providing reliable 
predictions. Fortunately, 1-dimensionality allows to exploit 
exact analysis techniques which can
provide --although only for some particular cases--
rigorous informations on the structure of the ground state and 
on low-energy excitations. For this reason a probative test 
for theoretical models is the comparison between experimental 
results and theoretical predictions on the ground state phase diagram.
\\The ground state is usually given as a function of the filling 
$n$, {\it i.e.} the number of effective carriers, and of a 
`band-parameter', which indicates the intrinsic unit of energy 
of the system (its actual definition depends on the theoretical 
approach envisaged, see below). One can thus distinguish between 
filling-controlled (FC) and band-controlled (BC) transitions, 
according to which kind of parameter (number of carriers 
or energy scale respectively) is tuned to let the transition occur.
Both kinds of transitions are very important in practical applications: 
BC transitions are relevant for instance in Vanadium oxides, where one can
modify the band-width through hydrostatic pressure on the sample; FC
transitions are frequent in perovskite-like materials such
as ${\mathrm{R}}_{1-x} {\mathrm{A}}_{x} {\mathrm{Ti O}}_3$ (R=rare-earth 
ion, and A=alkaline-earth-ion), as well as in hole-doped compounds like 
${\mathrm{La}}_{1-x} {\mathrm{Sr}}_{x} {\mathrm{Cu O}}_{2.5}$. In
order to describe these materials, at least as far as their low
energy excitations are concerned, where single band picture are often
reliable, the class of extended Hubbard models provides an interesting
starting point. For these models, which involve strong electronic 
correlations, the band-parameter is usually \cite{REV} taken to 
be the on-site Coulomb repulsion ($U$), instead of $w=4t$ 
($t$ = hopping amplitude), the latter being typical in mean-field
approaches. A number of exact results have been obtained in terms of $n$
and $U$. For the ordinary Hubbard model, 
a FC metal-insulator-metal transition has been shown to occur 
at half filling ($n=1$) for any $U>0$ ($U$ being the on-site 
Coulomb repulsion); on the
contrary, no BC transition takes place for $U>0$. 
More recently, some models (\cite{EKS} and \cite{AAS}) were solved in
which a BC Insulator-Superconductor Transition occurs at half-filling
at finite values of $U > 0$, while
the usual FC Metal-Insulator-Metal transition takes place for $n=1$
and $U>U_c$.
\\At the best of the authors' knowledge, no detailed investigation has 
been devoted to either of the following issues: for the BC transitions
it has not been pointed out yet what interaction terms are relevant to
tune the critical value $U_c$ at which the transition occurs: this is
quite important because $U_c$ can assume different values according to
the chemical structure of the material. Secondly, for the FC transitions,
all the above models provide an insulating state only at half filling;
on the contrary, doped materials exhibit an insulating phase
for a {\it finite} region of filling values.
In this letter we examine the above subjects providing the exact ground
state phase diagram of some 1-D electronic models. In 
particular, we obtain rigorous results which allow us to both discuss the 
dependence of $U_c$ of BC transitions on the Hamiltonian parameters, and 
to find a FC metal-insulator transition between two finite regions of the 
phase diagram.
\\We consider here the most general 1-band extended isotropic
Hubbard model preserving the total spin and number $N$ of electrons,
which reads
\end{multicols}
\begin{eqnarray}
&{\cal H}& = - \sum_{\langle{j},{k}\rangle,\sigma }[t-X (\hat{n}_{{j},
-\sigma}
+ \hat{n}_{{k},-\sigma})+\tilde X \hat{n}_{{j},-\sigma}
\hat{n}_{{k},-\sigma} ] c_{{j},\sigma}^\dagger
c_{{k}, \sigma} +  U \sum_{j} \hat{n}_{{j},\uparrow}\hat{n}_{{j},
\downarrow} + {V\over 2}\sum_{\langle{j},{k}\rangle} \hat{n}_{j} 
\hat{n}_{k} \label{EHM}  \\
&+& {W\over 2} \sum_{\langle{j},{k}\rangle,\sigma,\sigma' }
c_{{j},\sigma}^\dagger c_{{k},
\sigma'}^\dagger c_{{j},{\sigma}'} c_{{k}, \sigma}
+ Y\sum_{\langle{j},{k}\rangle } c_{{j},\uparrow}^\dagger
c_{{j},\downarrow}^\dagger  c_{{k},\downarrow} c_{{k},
\uparrow}
+ P \sum_{\langle{j},{k}\rangle } \hat{n}_{{j},\uparrow} \hat{n}_{{j},
\downarrow} \hat{n}_{k} + {Q\over 2} \sum_{\langle{j},{k}\rangle }
\hat{n}_{{j},\uparrow} \hat{n}_{{j}, \downarrow} \hat{n}_{{k},\uparrow}
\hat{n}_{{k}, \downarrow}  
\, , \nonumber
\end{eqnarray}
\begin{multicols}{2}
\noindent In (\ref{EHM}) $c_{{j},\sigma}^\dagger , c_{{j},\sigma} \,$
are fermionic creation and annihilation operators on a 1-dimensional 
chain with $L$ sites, $\sigma \in \{\uparrow, \downarrow \}$, $\, 
\hat{n}_{{j},\sigma} = c_{{j}, \sigma}^\dagger c_{{j},\sigma}$, 
$\hat{n}_{j} = \sum_\sigma \hat{n}_{{j},\sigma}$, and
$\langle {j} , \, {k} \rangle$ stands for neighboring sites. $t$
represents the hopping energy of the electrons (henceforth we set $t=1$), 
while the subsequent terms describe
their Coulomb interaction energy in a narrow band approximation: $U$
parametrizes the on-site repulsion, $V$ the neighboring
site charge interaction, $X$ the bond-charge interaction, $W$ the
exchange term, and $Y$ the pair-hopping  term. Moreover, 
additional many-body coupling terms have been included in
agreement with \cite{dBKS}: $\tilde X$ correlates hopping
with on-site occupation number, and $P$ and $Q$ describe three- and
four-electron interactions. In the following we shall identify the 4 
physical states $|\uparrow \rangle$, $|\downarrow \rangle$, 
$| 0 \rangle$ and $|\downarrow \uparrow \rangle$ 
at each lattice site with the canonical basis $e_{\alpha}$ of ${\Bbb C}^{ 4}$, 
and denote $n_{\uparrow}=N_{\uparrow}/L$; $n_{\downarrow}=N_{\downarrow}/L$; 
$n_{o}=N_{o}/L$; $n_{\uparrow \downarrow }=N_{\uparrow \downarrow}/L$  
the densities of these 4 species of physical states.
\\In \cite{DOMO1} it has been shown that, by fixing all the coupling 
constants of (\ref{EHM}) to appropriate values, one can rewrite ${\cal H}$
as a {\it generalized permutator} (GP) between neighboring sites (minus 
some constant terms). Here we add to the latter a further arbitrary
term $U \sum_j \hat{n}_{j \uparrow} \hat{n}_{j \downarrow}$, which is 
easily proved to commute with the GP. In matrix representation, the 
Hamiltonian (\ref{EHM}) that we consider reads
\begin{equation}
H = -  
\sum_{\alpha \ge \beta} \, \Pi_{\alpha \beta} \, + \,  
U N_{\uparrow \downarrow} \, - \,  const. terms \label{ham}
\end{equation}
where $\Pi_{\alpha \beta}$ acts as a GP $\Pi$ (see below) whenever 2 
neighboring sites of the chain are occupied by $e_{\alpha}$ and 
$e_{\beta}$, otherwise it gives zero. The constant terms are of the form 
${\bar U} N_{\uparrow \downarrow} + \bar{\mu} N + \bar{c} \, {\Bbb I}$,
where ${\bar U}, {\bar \mu}$ and ${\bar c}$ are fixed values. 
\\The purpose of this letter is to show how to investigate the ground 
state phase diagram of (\ref{ham}) as a function of the band parameter
$U$ and the filling of the carriers $n$.

Let us first recall some basic properties of the GP's. With respect 
to an ordinary permutator, a generalized permutator can either permute 
or leave unchanged the states of the 2 neighboring sites (including a 
possible additional sign); explicitly:
\be
\Pi \, (e_\alpha\otimes e_\beta) =
\theta^{d}_{\alpha \beta}  \,
(e_{\alpha} \otimes e_{\beta}) \, + \,
\theta^{o}_{\alpha \beta} \,
(e_{\beta} \otimes e_{\alpha}) \label{GP} 
\ee
where $\theta^{d}_{\alpha\beta}$ and $\theta^{o}_{\alpha\beta}$ are two 
discrete valued (0, -1 or~1) functions
determining on $\Pi$ the positions and the signs 
of the diagonal and off-diagonal entries respectively. Also, 
$\theta^{d}$ and $\theta^{o}$ are `complementary', {\it i.e.}
$|\theta^{d}_{\alpha\beta}|=1-|\theta^{o}_{\alpha\beta}|$,  
so that $\Pi$ has only one non-vanishing entry for row or column. 
Moreover, $\theta^{o}_{\alpha\beta}=\theta^{o}_{\beta\alpha}$, hence 
$\Pi$ is a symmetric matrix. The set of couples of subscripts
$(\alpha, \beta)$ for which $\theta^{d}_{\alpha \beta} \neq 0$ 
(resp.$\theta^{o}_{\alpha \beta} \neq 0$) is denoted by ${\cal A}^{d}$ 
(resp.${\cal A}^{o}$). 
It is easily seen that ${\cal A}^{d}$ is always of the form ${\cal A}^{d}=
\cup_i \, {\cal S}_i \times {\cal S}_i$, the ${\cal S}_i$'s being disjoint 
subsets of the set ${\cal S}=\{1,2,3,4 \}$. By varying the functions 
$\theta^{d}$ and $\theta^{o}$ one obtains different kinds of GP's. \\

In order to determine the ground state phase diagram of (\ref{ham}), 
the difficult task is to calculate the contribution to the energy of the 
first term. To solve this issue, one can reconduct it to a problem 
defined in a smaller Hilbert space. Indeed it can be seen that:\\

\noindent 1) Under the conditions precised below, a generalized 
permutator between physical states is equivalent to an ordinary 
permutator between the so called Sutherland's species (SS). 
The latter do not need to be identified with the physical species 
(PS) of the states. Indeed, the number of PS is determined by the nature 
of the problem (in our cases they are always the four $e_\alpha$),
while the number of SS is determined by the {\it structure} of the GP
entering the Hamiltonian. In particular, it may happen that different PS 
constitute a single SS, so that the number of the latter is 
$\le 4$, leading to a reduction of the dimensionality of the effective 
Hilbert space. Through a suitable identification of the Sutherland's 
species, the first term of (\ref{ham}) can be rewritten in the form:
\be
H_0 = -
\sum_A p_A N_{AA} - \sum_{A>B} \sigma_{AB} \Phi_{AB}  \quad , \label{HSUT}
\ee
where $p_A=\pm 1$ determines the nature of the $A$-th species, 
even (${\cal E}$)/odd (${\cal O}$) for $+1/-1$; $N_{AA}$ is the number of
neighboring sites occupied by the same species $A$, and
$\Phi_{AB}$ permutes objects of species $A$ and $B$ that occupy two
neighboring sites, otherwise it gives zero. The $\sigma_{AB}$ are signs. 
For a given GP, the SS are to be identified through the subsets
${\cal S}_i$'s of ${\cal A}^{d}$. In practice, the reduction to
Sutherland's species is possible if: $a)$ $\theta^{d}_{\alpha \beta}=p_{i}
\, \, \forall \alpha, \beta \in {\cal S}_i$ and $b)$ $\theta^{o}_{\alpha
\beta}=\sigma_{ij} \, , \, \forall \alpha \in {\cal S}_i \, , \, \forall
\beta \in {\cal S}_j$, where $i \neq j$.\\ 

\noindent 2) In the case where (\ref{HSUT}) has $\sigma_{AB}$=+1 
$\forall \, A,B \in {\cal E}$, 
and $\sigma_{AB}$ independent of 
$B$ for $B \in {\cal O}$ and $A \in {\cal E}$ (Sutherland's 
Hamiltonian), it is possible to reduce the number of 
even species down to only one. Indeed in this case the ground state
energy of hamiltonian (\ref{HSUT}) for a  system 
with $x$ even species and $y$ odd species is equal to that of the same 
hamiltonian acting on a system with the same number of odd species but 
just one even species collecting all the previous ones (as implied by a simple 
extension of Sutherland's theorem, see \cite{SUT}).\\

\noindent {\it Remark}: For a given GP, the fulfillment of the 
conditions given at points 1) and 2) depends also on the normalization 
chosen to define the basis vectors. It is worth emphasizing that some GP, 
though apparently violating the above requirements, can be brought to 
fulfill them through a mere redefinition of the phase of a given physical 
species $\bar{\alpha}$, {\it i.e.} $|e_{\bar{\alpha}}\rangle_j \rightarrow 
(-1)^{j} |e_{\bar{\alpha}} \rangle_j$. We shall make use of this remark 
in the following.\\

To illustrate how the above observations can be exploited, we start 
with a known case, the AAS model \cite{AAS}, which differs from the
ordinary Hubbard model only for a correlated-hopping term ($X=1;
\tilde{X}=V=W=Y=P=Q=0$). This model is of the form (\ref{ham}), with 
${\bar U}=-4$, $\bar{\mu}=2$, and $\bar{c}=-1$. Its GP reduces to
(\ref{HSUT}) by identifying the following two Sutherland's species:
$A=\{ |\uparrow\rangle ,|\downarrow \rangle \}$ (odd) and $B=\{ |0
\rangle , |\downarrow \uparrow \rangle \}$ (even). In this formalism
the model is nothing but a free spinless fermion model, and its energy
per site is given by  
\be
\epsilon=2 n_{A}-1- {2\over \pi}  \sin(\pi n_A) + (U-{\bar U}  )n_{\uparrow
\downarrow} -\bar{\mu} n -\bar{c} \, ,    \label{gse}
\ee
where $n_{A}=n_{\uparrow}+n_{\downarrow}$. The phase diagram as a function
of the filling $n=N/L$ and $U$ can be easily 
derived by exploiting the identity $n_{\uparrow \downarrow}=(n-n_A)/2$ and
minimizing $\epsilon$ with respect to $n_A$ ($n_A\in [0,n]$ for 
$0 \le n \le 1$ and $n_A\in [0,2-n]$ for $1 \le n \le 2$) and
coincides with that derived in \cite{AAS}. We also notice that the model
with $X=1; \tilde{X}=2; V=W=Y=P=Q=0$ has the same energy as
AAS. In fact, using the above {\it Remark} and redefining the basis
vector $|\downarrow \uparrow \rangle_j \rightarrow (-1)^j |\downarrow
\uparrow \rangle_j$, it can 
be cast in the form (\ref{ham}) with the same GP as AAS.

\begin{figure}
\epsfig{file=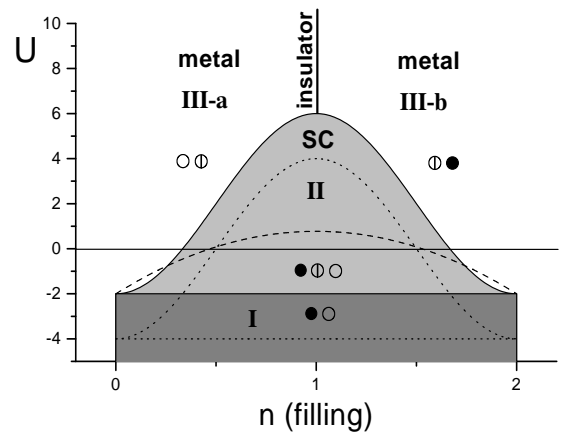,width=8cm,height=12cm}
\caption{Ground state phase diagram of the model $X=1; \tilde{X}=(1-\sigma); 
Y=-\sigma; P=-1; Q=2$. The model exhibits a BC transition 
insulator-superconductor transition at $n=1$, for $U_c=6$. The dashed line is 
the EKS model, and the dotted line corresponds to the AAS model. }
\label{sut_fig1}
\end{figure}

The method just outlined allows the solution of a wide class of 
models whose Hamiltonian has the form (\ref{ham})\cite{DOMO1}. In
particular, as the AAS model displays both a BC and a FC
transitions, here we apply it to the study of similar, but new, models
in which further terms are included, yielding a change in the value of
the parameters characterizing the transition. We consider the model 
with coupling constants $X=1; \tilde{X}=(1-\sigma); Y=-\sigma ;
P=-1 ; Q=2$ where $\sigma=\pm 1$. The resulting Hamiltonian has the form
(\ref{ham}) with ${\bar U}=-2$, $\bar{\mu}=2$, $\bar{c}=-1$ in both
cases $\sigma = \pm 1$. $\Pi$ has
diagonal entries characterized by the subsets ${\cal S}_1=\{ 1,2 \}$ ,
${\cal S}_2=\{ 3 \}$ , ${\cal S}_3=\{ 4 \}$, and off-diagonal entries
$\theta^{o}_{\alpha \beta} = +1$ $\forall \alpha \in {\cal S}_1 \, , \,
\beta \in {\cal S}_2$ and $\theta^{o}_{\alpha \beta} = \sigma$ $\forall
\alpha \in {\cal S}_1$ or ${\cal S}_2 \, , \,  \beta \in {\cal S}_3$. 
Both the conditions $a)$ and $b)$ to identify Sutherland's species are
thus fulfilled, and the species read: $A=\{ |\uparrow \rangle , 
|\downarrow \rangle \}$ (which is `odd' because $\theta^{d}_{\alpha
\beta}=-1$ if $\alpha,\beta \in {\cal S}_1$) ; $B=|0 \rangle$ (`even'
because $\theta^{d}_{33}=+1$) and $C=|\downarrow \uparrow \rangle$
(`even' because $\theta^{d}_{44}=+1$). For the case $\sigma=+1$ one can
straightforwardly apply Sutherland's theorem (see point 2)) to reduce
the number of even species to 1, ending up with a free spinless
fermion problem, where occupied sites are represented by $A$ and 
empty sites by $B$ and $C$. The ground state energy $\epsilon$ per 
site has the same form as (\ref{gse}), in which again 
$n_{\uparrow\downarrow}=(n-n_A)/2$, and it has to be minimized 
with respect to $n_A$. For the case $\sigma=-1$,
before using Sutherland's theorem, one has to apply again the
{\it Remark}, changing $|\downarrow \uparrow \rangle_j
\rightarrow (-1)^j |\downarrow \uparrow \rangle_j$. The expression of
$\epsilon$ is identical.
\\The phase diagram is given in fig.\ref{sut_fig1}; the lower region I
is characterized by $n_{A}=0$, so that only doubly occupied or empty
sites are present in the ground state; in this region the ground state
$|\Psi_0 \rangle$ is made of the so-called (pure) {\it eta-pairs},
{\it i.e.} $|\Psi_0 \rangle=(\eta^{\dagger}_{\varphi})^{N/2} |0 \rangle $,
where $\eta^{\dagger}_{\varphi}=\sum_j e^{i \varphi j}
c^{\dagger}_{j \uparrow} c^{\dagger}_{j \downarrow}=
\sum_{k} c^{\dagger}_{k \uparrow} c^{\dagger}_{\varphi-k \downarrow}$
with pair momentum $\varphi=0,\pi$. In the case $\sigma=+1$ we have 0-pairs,
whereas if $\sigma=-1$ the pairs have $\pi$-momentum. The latter case is
particularly important because the $\pi$-pairs (and not $0$-pairs) are
expected to survive as the constraint $X=1$ is relaxed (see \cite{MOCA}).
In region II, delimited by $U_{\mbox{\tiny II-III}}=
2-4\cos (\pi n)$, we have the simultaneous presence of 
empty (\raisebox{-.5ex}{\LARGE{$\circ$}}), 
singly occupied (\raisebox{-.52ex}{\LARGE{$\circ$}} 
$\hspace{-0.9em} \mathbf{\shortmid} \hspace{0.42em}$) 
and doubly occupied (\raisebox{-.5ex}{\LARGE{$\bullet$}}) sites; this
is called {\it mixed} region and the ground state is $|\Psi_0 \rangle=
(\eta^{\dagger}_{\varphi})^{N/2} |U=\infty \rangle $, where 
$|U=\infty \rangle$
are the eigenstates of the $U=\infty$ Hubbard model\cite{dBKS}.
In both region I and II the ground state is superconducting, because 
the 2-particle reduced density matrix exhibits long range
correlation\cite{AAS}, {\it i.e.} 
$g(i,j)=\langle \Psi_0|c^{\dagger}_{i\uparrow} 
c^{\dagger}_{i \downarrow} c^{}_{j \downarrow} c^{}_{j \uparrow} 
|\Psi_0 \rangle \, \nrightarrow 0$ for $|i-j| \rightarrow + \infty$.
Finally, the region III-a ($0 \le n \le 1$) is made of singly occupied and 
empty sites; in this region the ground-state of the $U=\infty$ Hubbard
model is eigenstate of the Hamiltonian and is metallic. The region III-b
($1 \le n \le 2$) is the particle-hole transformed of III-a, and the
metallic carriers are holes. One can show that at half-filling the system
is an insulator with gap $\Delta=U-6$. 
\\With respect to the AAS model we observe that the pair-hopping 
term has two main effects: first it removes the degeneracy in 
$\varphi$ in this region (only $\varphi=0$ or $\pi$ survive, according 
to the sign $\sigma$ of $Y$); secondly it raises the borderline of 
such region upwards: in fact it can be generally shown 
\cite{POL} that a pair hopping term acts as an effective attraction  
($\propto \, -|Y|$) renormalizing the Coulomb repulsion $U$. The 
superconducting region II of our model is enhanced also with respect 
to that of the EKS model. Indeed, although the pair-hopping term is 
also present in the EKS model (the borderlines of region I coincide), 
its effect is strongly reduced near half-filling due to the Coulomb attraction
term between neighboring sites ($V=-1$), which is known to compete with
the formation of on-site pairs. 
\\As a consequence, the BC insulator-superconductor transition occurring
at half-filling corresponds to the maximum critical value
$U_c^{max}=6$, higher than for all other exactly solved models. This is
important because higher values of $U_c$ reduce the probability that
thermal fluctuations may destroy the superconducting phase.\\

\begin{figure}
\epsfig{file=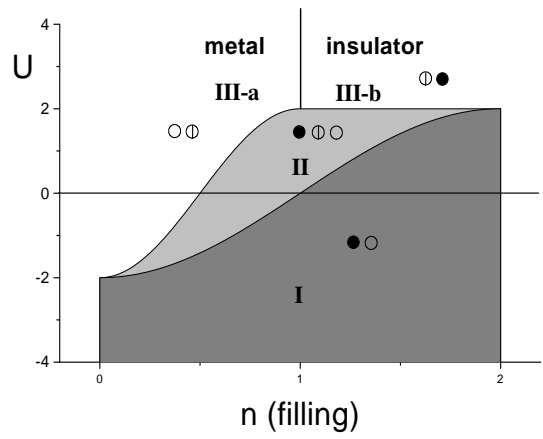,width=8cm,height=12cm}
\caption{Ground state phase diagram of the model $X=\tilde{X}=1$ ; 
$Y=\sigma$; $V=W=P=Q=0$. A FC metal-insulator transition takes place 
for $U \ge 2$  between two {\it finite} regions, III-a and III-b.}
\label{sut_fig2}
\end{figure}

Because of the particle-hole symmetry of the models we have considered
so far, the insulating phase can exist just at half filling. In order
to investigate FC transitions between {\it finite} metal-insulator regions
of the phase diagram, we now discuss a simple model not particle-hole
invariant, describing a competition between the $U=\infty$ Hubbard model
(excluding doubly-occupancy), and the pair-hopping (favouring the
formation of pairs), modulated by the band parameter $U$ (explicitly
$X=\tilde{X}=1$; $Y=\sigma$; $V=W=P=Q=0$). It is easy to realize that
(up to the application of the {\it Remark}) this model can be set in the
form (\ref{ham}) (${\bar U}=-2$, $\bar{\mu}=2$,
and $\bar{c}=-1$). The GP is now equivalent to an ordinary permutation
between the two Sutherland's species: $A=\{ |\uparrow \rangle ,
|\downarrow \rangle,
|\downarrow \uparrow \rangle\}$ (which is `odd' because 
$\theta^{d}_{\alpha \beta}=-1$ if $\alpha,\beta \in {\cal S}_1=\{1,2,4\}
$); $B=|0 \rangle$ (`even' because $\theta^{d}_{33}=+1$). 
The ground state energy per site is still given by eq. (\ref{gse}),
where now $n_{\uparrow \downarrow}=n-n_A$, . The phase diagram
--obtained by minimizing $\epsilon$ at fixed $n$ with respect to $n_A$
in the range $n/2\leq n_A\leq \min (n,1)$-- is presented in fig.
\ref{sut_fig2}, and exhibits again four regions. However, due to the
absence of particle-hole invariance, the shape is not symmetric around
half-filling.
\\In region I (just doubly occupied and empty sites) only the $Y$ and
$U$ terms act: the model behaves like a spin isotropic XX model
($\tilde{S}^{+}_i=c^{\dagger}_{i \uparrow} c^{\dagger}_{i \downarrow}$,
$\tilde{S}^{-}_i= c^{}_{i \downarrow} c^{}_{i \uparrow}$) with the $U$
term acting as a magnetic field; it is well known that at $U=0$ the
correlation function has a power law decay $g(i,j) \propto |i-j|^{-1/2}$,
whereas $g$ is not known for non-vanishing magnetic field. However, as
far as $U\leq 2$, long-range order arises for any non-zero value of
anisotropy. The borderline of this region is given by 
$U_{\mbox{\tiny I-II}}=-2 \cos(\pi n/2)$. Notice that this 
region raises up to {\it positive values} of $U$ for $1 \le n \le 2$.
The mixed region II is entered as the double occupancy begins to
decrease from its maximum value, yielding the increase of the local
magnetic moment $M_0 = 3/4 \, L^{-1} \sum_j \langle \Psi_{0}|
(\hat n_{j,\uparrow}-\hat n_{j,\downarrow})^2 |\Psi_{0} \rangle=$
$3/4 \,( 2\pi^{-1} 
\arccos(-U/2)-n)$. The value of $n_{\uparrow \downarrow}$ reaches 
its minimum
for $U_{\mbox{\tiny II-IIIa}}= -2 \cos (\pi n)$ when $n\leq 1$, and for
$U_{\mbox{\tiny II-IIIb}}=2$ when $n\geq 1$. Correspondingly, regions
III-a and III-b are entered.
The former is metallic, the ground-state is that of the
$U=\infty$ Hubbard model, and the system behaves like a Tomonaga-Luttinger
liquid. The most interesting feature is that region III-b is a
{\it finite} insulating region. More precisely, at exactly
half filling the gap is $\Delta=U-2$, while for $1 < n \leq 2$ 
no empty site is present, and the model behaves like the Hubbard
model in the atomic limit. Hence here the FC
transition takes place between two finite regions, in analogy with
experimental observations on chain hole-doped compounds. 
Interestingly, for the the special value $U=2$, our model and the 
$U$-supersymmetric model coincide. As a consequence, our
ground state energy in this case is equal to that obtained
in~\cite{BEFR}.

In this letter we have presented exact ground state phase diagrams of
two electron models, and studied their BC and FC transitions. Our
analysis supports the relevance of the pair-hopping term in raising the
critical value of $U$ for BC superconducting-insulator transitions, as
well as the importance of particle-hole not invariant terms in the
apperance of a finite insulating region. The method we used can be
implemented on all those models described by Hamiltonian 
(\ref{ham}) in which the GP
verifies conditions {\it a}) and {\it b}).  We stress that such
GPs all correspond to integrable models \cite{DOMO1}, {\it i.e.}
they are solutions of the Yang-Baxter Equation (consistency equation
for factorizability). The Hamiltonians exhibit therefore a set of
conserved quantities mutually commuting. 


\end{multicols}

\begin{references}
\bibitem{REF} M.Dumm {\it et al.}, Phys. Rev. B {\bf 61}, 511 (2000); 
R.~Farchioni, P.~Vignolo, and G.~Grosso, Phys. Rev. B {\bf 60}, 15705 (1999); 
E.~Chow, P.~Delsing and D.B.~Haviland, Phys. Rev. Lett. {\bf 81}, 204 (1998) 
\bibitem{REV} M. Imada, A. Fujimori, Y. Tokura, Rev. Mod. Phys. 
{\bf 70}, 1039 (1998)
\bibitem{EKS} F.H. Essler, V. Korepin, and K. Schoutens, Phys. Rev.
Lett. {\bf 68}, 2960 (1992); Phys. Rev. Lett. {\bf 70}, 73 (1993)
\bibitem{AAS} L.Arrachea, and A.Aligia, Phys. Rev. Lett. {\bf 73},
2240 (1994); A.Schadschneider, Phys. Rev. B {\bf 51}, 10386 (1995)
\bibitem{dBKS} J.~de Boer, V.~Korepin, A.~Schadschneider, Phys. Rev.
Lett. {\bf 74}, 789 (1995); 
C.~Castellani, C.~Di Castro, M.~Grilli, Phys. Rev. Lett. {\bf 72}, 
3626 (1994)
\bibitem{DOMO1} F. Dolcini, and A. Montorsi, Int. J. Mod. Phys. B 
(to be published)
\bibitem{SUT} B. Sutherland, Phys. Rev. B {\bf 12}, 3795 (1975)
\bibitem{MOCA} A. Montorsi, and D.K. Campbell, Phys. Rev. B {\bf 53},
5153 (1996) 
\bibitem{POL} S. Robaszkiewicz, B.Bulka, Phys. Rev. B 
{\bf 59}, 6430 (1999) 
\bibitem{BEFR} G.Bed\"{u}rftig, H. Frahm, J.Phys. A; {\bf 28} 4453 (1995)
\end{references}
\end{document}